\documentclass[intlimits,twoside,a4paper]{article}
\usepackage{amsmath,amssymb}
\usepackage{graphicx}

\usepackage[T2A]{fontenc}
\usepackage[cp1251]{inputenc}

\usepackage[eqsecnum]{cmpj2}
\issue{2013}{16}{3}{33803}
\doinumber{10.5488/CMP.16.33803}

\title[Numerical analysis of the TiN/Al$_{2}$O$_{3}$ coating]
{Numerical analysis of the morphological and phase changes in the TiN/Al$_{2}$O$_{3}$ coating under high current electron
beam modification}
\author[A.D. Pogrebnjak, V.N. Borisyuk, A.A. Bagdasaryan]
{A.D. Pogrebnjak, V.N. Borisyuk, A.A. Bagdasaryan}
\address{Sumy State University, 2 Rimskii-Korsakov St., 40007 Sumy, Ukraine}

\date{Received May 7, 2013, in final form July 10, 2013}
\authorcopyright{A.D. Pogrebnjak, V.N. Borisyuk, A.A Bagdasaryan, 2013}

\begin{document}

\maketitle

\begin{abstract}
A modification of the surface structure of the hybrid coating
TiN/Al$_{2}$O$_{3}$ with a low-energy high-current electron beam
(NCEB) is performed. The surface roughness is considered as a
function of beam current. Surfaces of the obtained samples are
investigated within the two-dimensional multifractal detrended
fluctuation analysis (MF-DFA). The multifractal spectrum of the
surface is calculated as a quantitative parameter of the roughness.
It is shown that with an increase of the beam energy, the surface
becomes more regular and uniform.
\keywords self-similarity, fractal dimension, hybrid coating, high-current electron
beam(s) effect
\pacs 81.40.- z, 81.07.- b, 68.35.-p, 61.46.-w, 05.45.Df, 61.43.Hv

\end{abstract}

\section{Introduction}

Many objects and systems in nature exhibit a self-similar or
self-affine structure \cite{Fed98, Ole09}. Self-similarity means that
each segment of the initial set has the same structure as the whole
object.The properties of such structures can be described by
specific parameters, such as fractal dimension (or set of dimensions in
the case of the multifractal objects \cite{Ol95}), Hurst exponent
and others. Common examples of such objects are the Koch curve and
the Cantor set. The property of self-similarity is inherent not only
to topological structures but also to the phase space of complex
stochastic systems with a hierarchical structure and non participant
regions. The stochastic fractals can be illustrated through the
Lorenz attractor and  non-stationary time series \cite{Ol09}.

The surface roughness characterization is an important problem for
both applied and theoretical science. Many image techniques have
been extensively used to investigate the rough surfaces, such
as atomic force microscopy, secondary electron microscopy, optical
imaging techniques and others \cite{Arn00}. It has been established
\cite{Jen03} that the roughness parameters based on conventional
theories depend on the sampling interval of a particular measuring
instrument used. Using the methods of fractal geometry, this problem
is eliminated because the fractal model includes topological
parameters that do not depend on the resolution of the instrument
used. The concept of fractal dimension, in contrast to traditional
methods, has made it possible to explain the physical properties of
the system depending on the geometry of the surface
\cite{Pfe89,Bor20}. For this purpose, many methods have been
proposed, such as detrended fluctuation analysis (DFA) \cite{Gu06},
two-dimensional multifractal detrended fluctuation analysis (2D
MF-FDA) \cite{Li09} a generalization of the 1D DFA and MF-DFA
\cite{Niu08}, rescaled range analysis \cite{Hur51} and many more
\cite{Ol20}.

In the present article we present the investigation of a self-similar
structure of the TiN/Al$_{2}$O$_{3}$ hybrid coating surfaces using
numerical methods of scaling analysis. Our calculations are based on the
algorithm of two-dimensional multifractal detrended fluctuation
analysis (MF-DFA) \cite{Gu06}. This algorithm was initially was
developed for investigation of the time series as a one-dimensional
self-similar set \cite{Kan02}, and later on generalized for the analysis of more
complicated objects \cite{Gu06}. Our calculations make it possible to
present a quantitative characteristic of the surface roughness, and
to compare it for different samples. Earlier in the work \cite {Bag12}
we have considered the dependence of the generalized
Hurst exponent surface of the coating TiN/Al$_2$O$_3$ at different beam
current densities. However, a complete representation of multifractal
formalism is achieved by calculating the multifractal spectrum $f(\alpha)$ and
singularity strength $\alpha$. This knowledge would be useful
because the roughness of hybrid coatings is an important factor at contact wear and
physical phenomena such as absorption, catalysis and the dissolution
of a fractal object.

The paper is organized as follows. In the opening section we briefly
describe the object of our research~--- TiN/Al$_{2}$O$_{3}$ hybrid
coatings, obtaining through plasma-detonation the technology and
additional treatment by high-current electron beams (HCEB) at
different regimes of partial melting. Next, we refer the main steps
of the MF-DFA algorithm and present the results of our calculations,
comparing them for different samples. The final section is devoted
to discussion.

\section{Samples under investigation}

We used the
$\alpha$-Al$_{2}$O$_{3}$ powder with a particle size of 27 to 56
microns as initial material for the deposition, which was applied in the facility ``Impulse-5'' on the
substrate of austenite steel AISI 321 (18 wt.\% Cr; 9 wt.\% Ni; 1
wt.\% Ti; 0.3 wt.\% Cr; Fe the rest; 0.3~mm and 2~mm thickness)
\cite{Pog06,Pog12,Pog20}. An oxide-aluminum ceramics and other
coatings based on titanium carbide and tungsten carbide and nitrides
possess a number of useful properties, which are capable of providing
a corrosion protection, high hardness and mechanical strength, low
wear and good electro-isolation properties. However, these coatings
are characterized by the presence of macro-, micro- and
submicroscopic porosity, and a certain number of defects
\cite{Kun07}. For this purpose, to increase the corrosion resistance
of protective ceramic coatings and to reduce the concentration of
defects caused by deposition, the surface was
coated with TiN layer. The deposition continued for 20~min in the
atmosphere of ionized nitrogen, under 700~K temperature, and about
$10^{-1}$ to $10^{-2}$ operation pressure of the reaction gas.
The deposition TiN layer was made by means of the facility ``Bulat-3T'' with a
vacuum-arc source (Kyiv, Ukraine).
\begin{figure}[!h]
\centerline{
\includegraphics[width=0.48\textwidth]{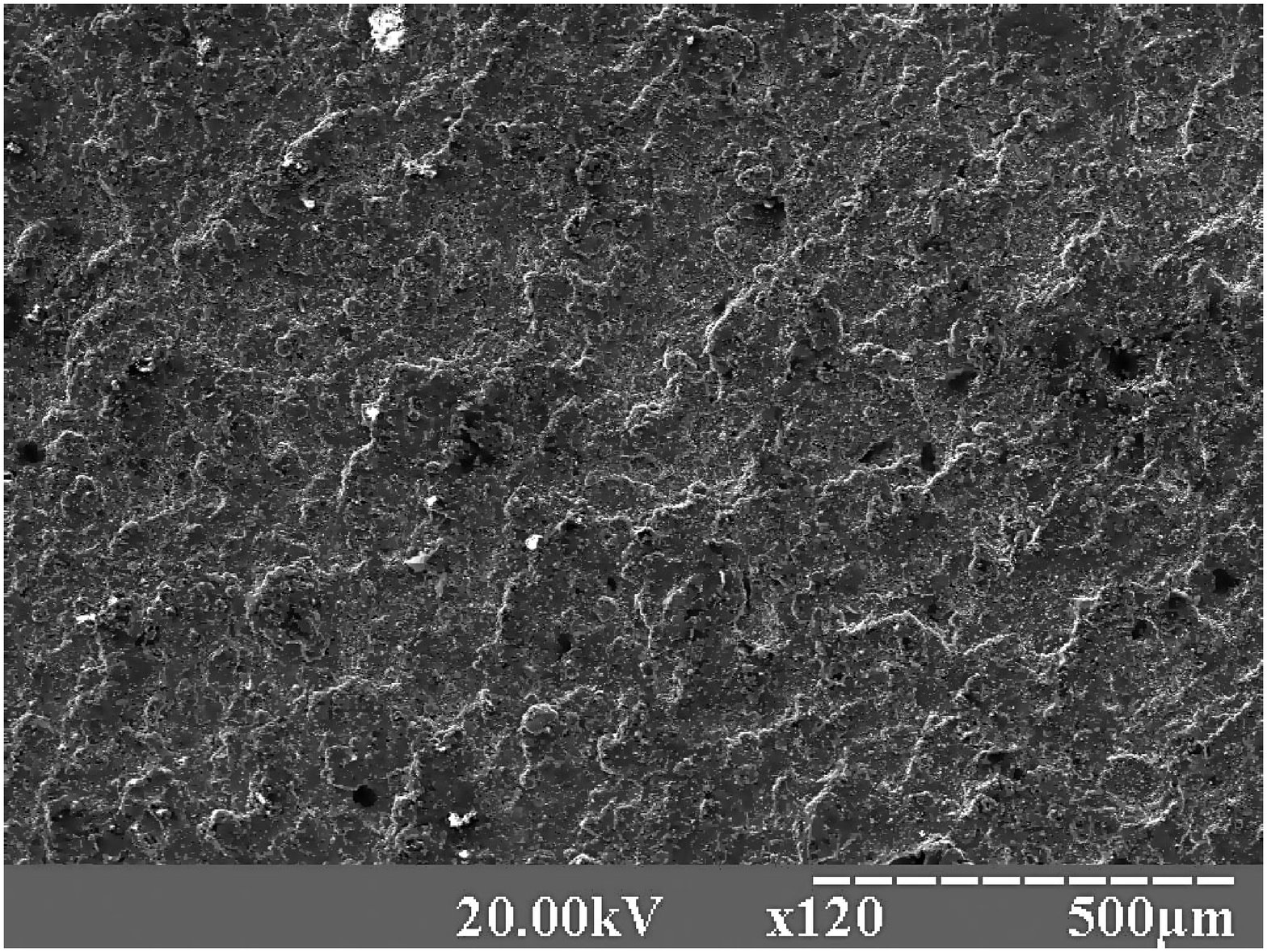}
\hfill
\includegraphics[width=0.48\textwidth]{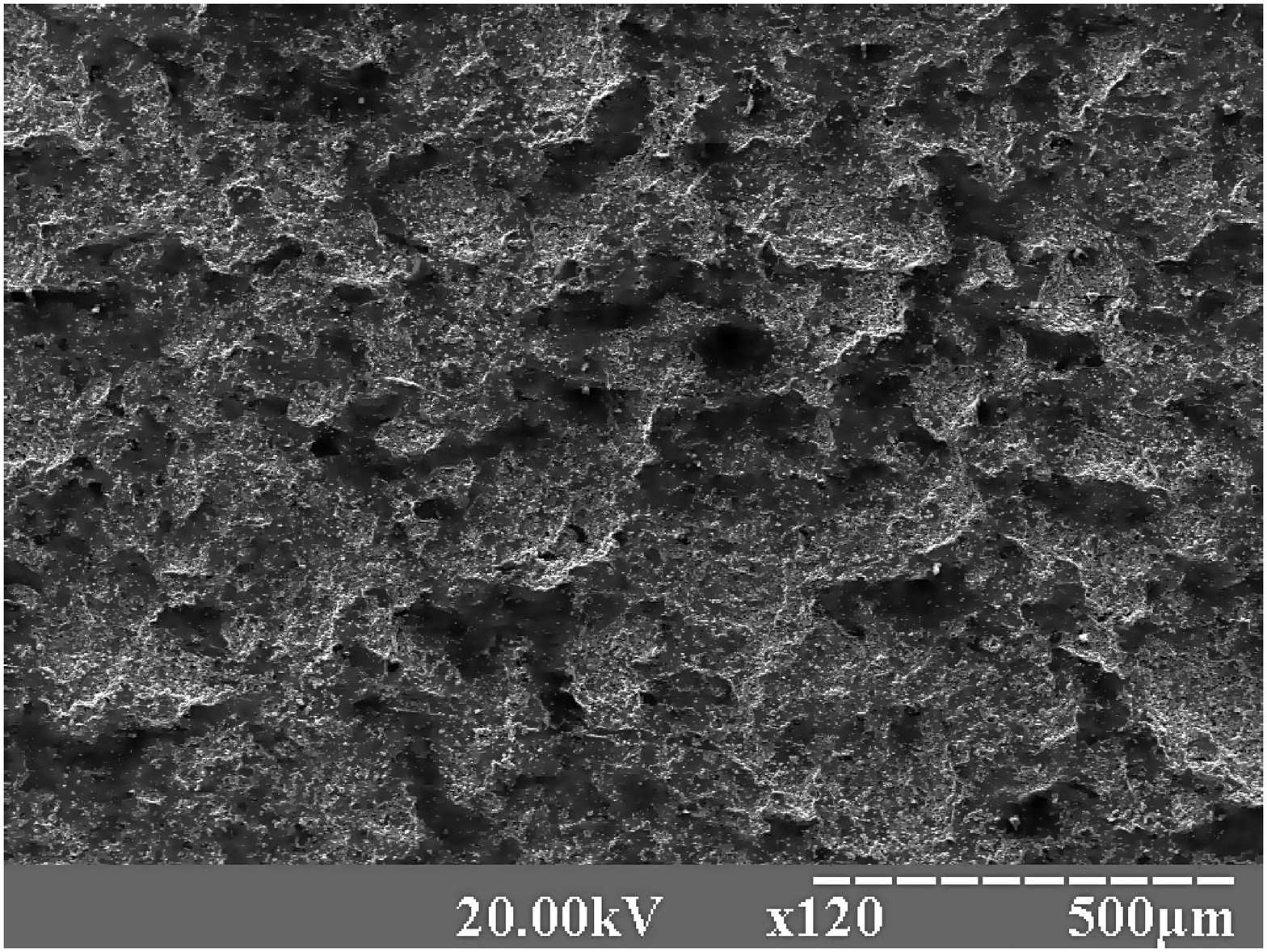}
}
\centerline{ (a) \hspace{55mm} (b) }
 \centerline{
\includegraphics[width=0.48\textwidth]{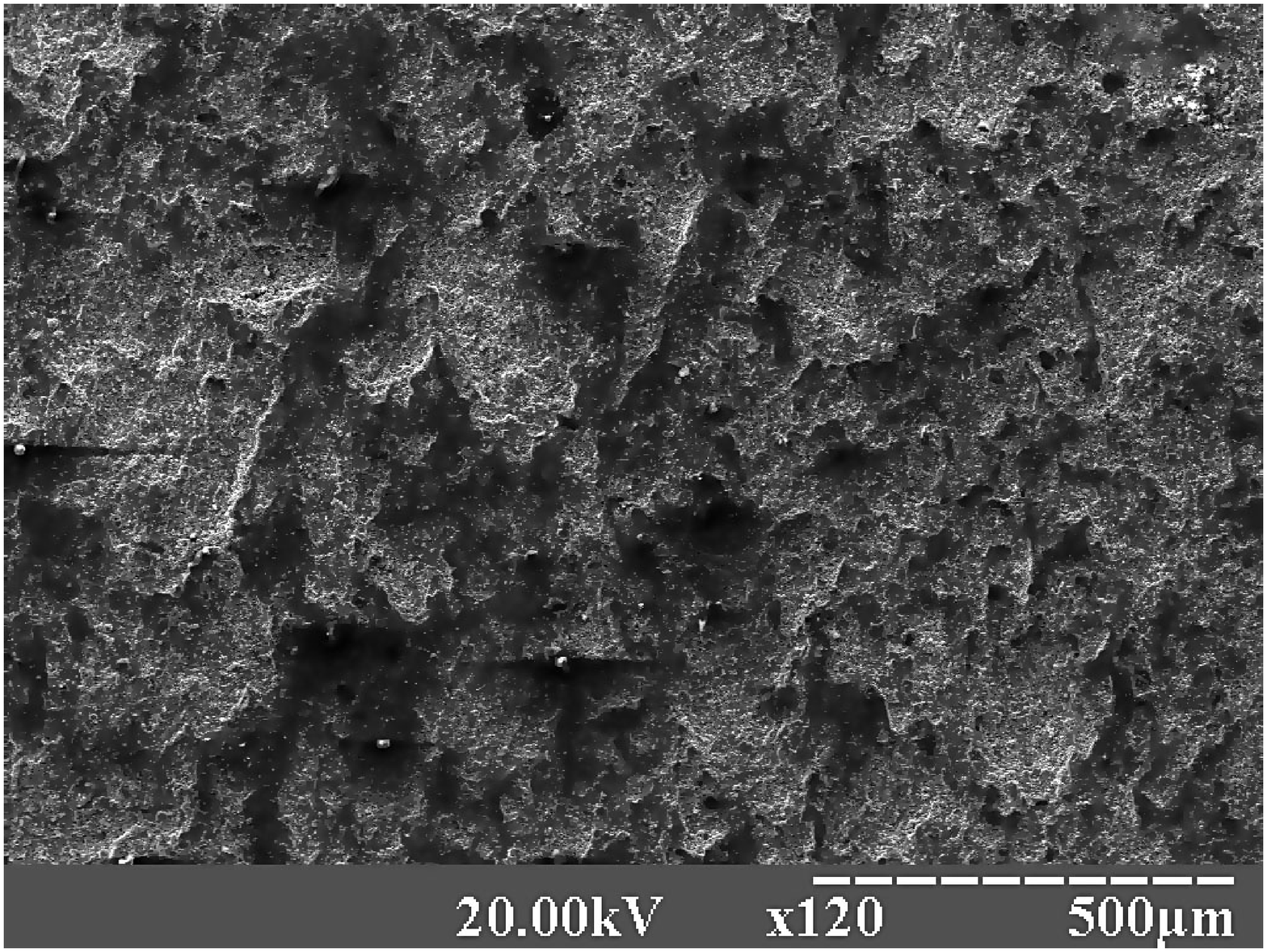}
\hfill
\includegraphics[width=0.48\textwidth]{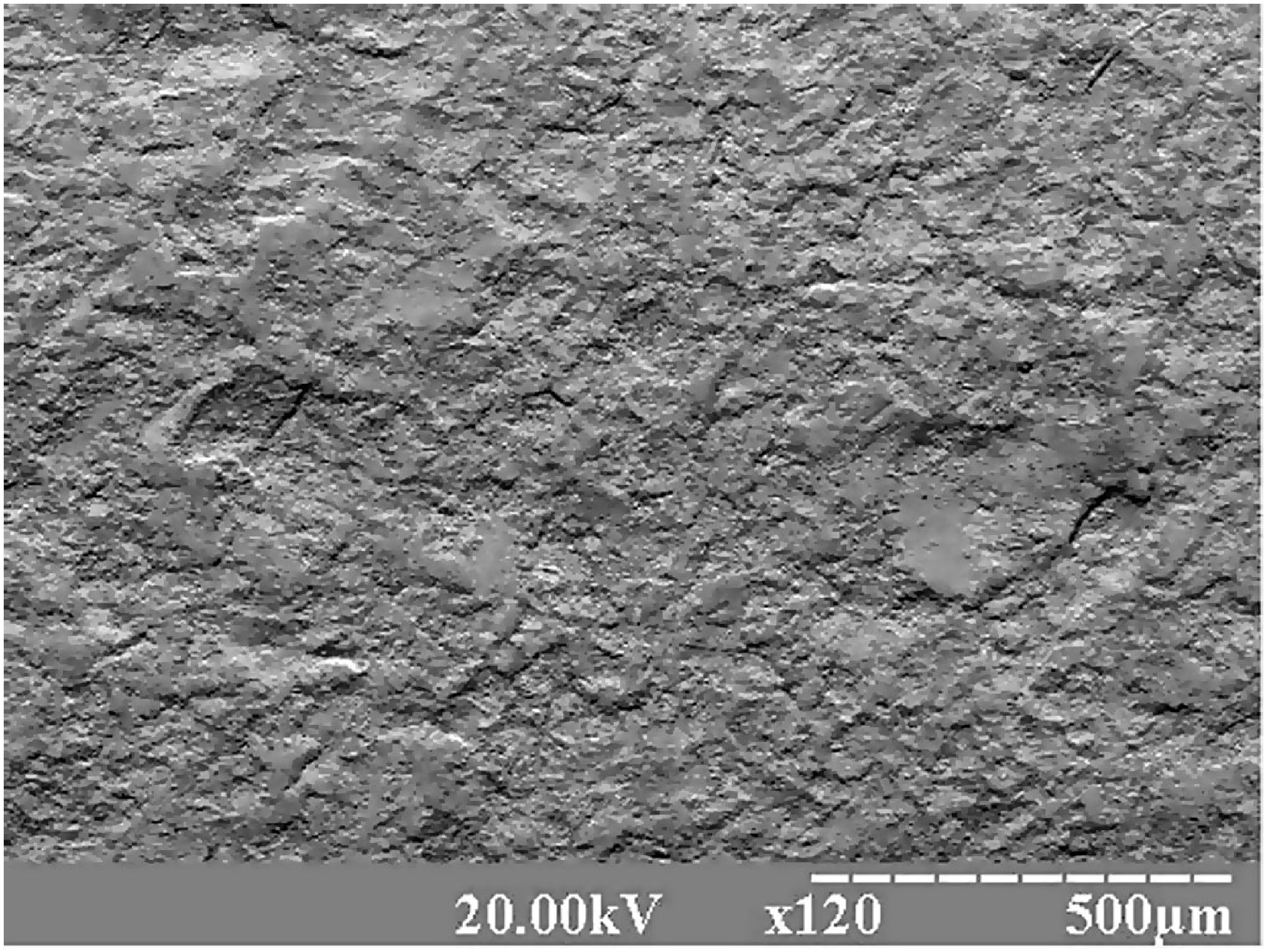}
}
\centerline{ (c) \hspace{55mm} (d) } \caption{SEM images of surfaces of
hybrid TiN/Al$_2$O$_3$ coatings under HCEB modification. The beam
current in the electron beam quenching of coating surfaces was,
respectively: (a)~--- $20$~mA, (b)~--- $20+15$~mA, (c)~--- $20+25$~mA, (d)~--- 20+35~mA.} \label{fig-1}
\end{figure}

One of the promising methods in solving the problem of adhesion of
thin film coatings and reducing the roughness of the powder sublayer is a
thermal treatment of the surface by high-current electron beams
(HCEB) in the regime of partial melting. This technology made it possible to
heal the micropores and to stimulate the diffusion processes between
the deposited particles and layers.

It was found that the electron beam melting of hybrid coating surfaces
TiN/Al$_2$O$_3$ (20~mA beam current) was accompanied by a partial
melting of non-uniformities occurring in the surface structure
(fi\-gu\-re~\ref{fig-1}). It can be seen that the coating had a layered but
melted structure. Regions of pit destruction were found on the
surface (dark points seen in the photo). These craters appeared as a
result of degassing induced by the electron beam melting of the
surface layers. In addition, there were found light-color
inclusions in the coatings. Repeated HCEB meltings of the coatings induced essential
(even visible) changes in the surface relief.

During the second stage of melting, the geometry the
surface layers of hybrid coatings  depended on the electron beam power density.
Correspondingly, the higher it was, the better these hybrid coating
surfaces mixed and the more uniform they became. The thermal
activation by the electron beam of coatings with the magnitude of 35~mA
is accompanied by intense changes in the geometry of the surface
layer. A complete fusion of the material near the surface is
observed. The coating has a developed structure and represents
a smooth alternation of peaks and valleys into each other, which is
characterized by an appreciable decrease in the surface roughness.

\section{Image analysis methodology}

All surfaces were investigated within two-dimensional multifractal
detrended fluctuation analyses (MF-DFA) methodology \cite{Kan02,
Gu06}. This algorithm allows one to calculate the main parameters of
the self-similar structure \cite{Fed98}.

Self-similar surface is considered as a two-dimensional data array
$X(i,j)$, where $i,j$ has discrete va\-lu\-es $i=1,2,\ldots,M$ and
$j=1,2,\ldots,N$. $X(i,j)$ itself is a surface obtained from the
digital electronic microscopy image by decomposing it according to
pixel indexes $(i, j)$ and brightness level $X$. This surface is
fragmented into $M_{s}\times N_{s}$ non-overlapping segments of the
sizes $s$, where $M_{s}=[M/s]$ and $N_{s}=[N/s]$ are integer
numbers.

For each segment $X_{\vartheta, \omega}$ identified by $\vartheta,
\omega$ the cumulative sum $u_{\vartheta\omega}(i,j)$:
\begin{equation}
\label{eq1} u_{\vartheta\omega}(i,j) = \sum^{i}_{k_{1}=1}
           \sum^{j}_{k_{2}=1}
           X_{\vartheta\omega}(k_{1}, k_{2})
\end{equation}
is calculated for all segments $\vartheta, \omega$.
The next step is a
detrending procedure for the obtained surface
$u_{\vartheta\omega}(i,j)$. The trend can be removed by the fitting
procedure, when it is determined by some smooth polynomial function
$\tilde{u}_{\vartheta\omega}(i,j)$. There are many possible
expressions of the $\tilde{u}_{\vartheta\omega}(i,j)$ function, but we
choose the simplest one, in order to reduce the computational time:
\begin{equation}
\label{eq2} \tilde{u}_{\vartheta\omega}(i,j) = ai+bj+c,
\end{equation}
where $a$, $b$, $c$ are coefficients defined by least-squares fitting
algorithm. It should be mentioned that more complicated forms of
$\tilde{u}_{\vartheta\omega}(i,j)$ do not provide any significant
improvement of the precision of the method, but noticeably increase
computational time \cite {Gu06}. After detrending we arrive at
a residual function:
\begin{equation}
\label{eq3} \varepsilon_{\vartheta\omega}(i,j) =
           u_{\vartheta\omega}(i,j)-
           \tilde{u}_{\vartheta\omega}(i,j),
\end{equation}
and a dispersion of the $\vartheta,\omega$ segment of length $s$:
\begin{equation}
\label{eq4} F^{2}(\vartheta,\omega,s)= \frac{1}{s^{2}}
            \sum^{s}_{i=1} \sum^{s}_{j=1}
            \varepsilon^{2}_{\vartheta\omega}(i,j).
\end{equation}
Dispersion of  all the segments is calculated through averaging over
all surfaces:
\begin{equation}
\label{eq5} F_{q}(s)= \left\{\frac{1}{M_{s}N_{s}}
            \sum^{M_{s}}_{\vartheta=1} \sum^{N_{s}}_{\omega=1}
            \left[F(\vartheta,\omega,s)\right]^{q}\right\}^{1/q},
\end{equation}
where $q$ is the deformation parameter initialized to increase the
role of the segments with small (when $q<0$) or high ($q>0$)
fluctuations $F^{2}(\vartheta,\omega,s)$, respectively. At $q=0$,
equation (\ref{eq5}) takes the form (\ref{eq6}):
\begin{equation}
\label{eq6} F_{0}(s)= \left\{\frac{1}{M_{s}N_{s}}
            \sum^{M_{s}}_{\vartheta=1} \sum^{N_{s}}_{\omega=1}
            \ln[F(\vartheta,\omega,s)]\right\}
\end{equation}
according to l'H\^opital's rule. For statistically correct
results, the values must be varied within the range from $s_\textrm{min}=6$ to
$s_\textrm{max}=\textrm{min}(M,N)/s$. The dispersion (\ref{eq5}) and the segment
size $s$ are linked through the scaling relation:
\begin{equation}
\label{eq7} F_{q}(s)\sim s^{h(q)},
\end{equation}
where $h(q)$ is the generalized Hurst exponent.

Equation (\ref{eq7}) can be rewritten according to the standard
multifractal formalism through scaling exponent $\tau(q)$ and
partition function $Z_{q}(s)$ as \cite{Kan02}:
\begin{equation}
\label{eq8}  Z_{q}(s)\sim s^{\tau(q)},
\end{equation}
\begin{equation}
\label{eq9}  Z_{q}(s)= \frac{1}{MN}
            \sum^{M/s}_{\vartheta=1} \sum^{N/s}_{\omega=1}
            [F(\vartheta,\omega,s)]^{q}.
\end{equation}

One can relate the H\"older exponent $\alpha$ and the
multifractal spectrum $f(\alpha)$ via Legendre transform
\cite{Pei92, Hal86}, deriving these multifractal parameters as
\begin{equation}
\label{eq10} \alpha=\tau'(q),
\end{equation}
\begin{equation}
\label{eq11} f(\alpha)=q\alpha-\tau(q).
\end{equation}
For monofractal objects, the function $\tau(q)$ is a linear dependence
which, with the transition to the multifractal, becomes more curved,
keeping the linear sections within $q\rightarrow\infty$. In the analyzed structure,
multifractality can be revealed more clearly
from the shape of the multifractal spectrum $f(\alpha)$, the width
of which provides a set of fractal dimensions (for example, for
monofractal curve $f(\alpha)$ has a $\delta$-function with the fixed
$\alpha$ value).

\section{ Multifractal analysis of experimental results}

In this section we apply the MF-DFA method to analyze the structure
of the surface of the hybrid coating TiN/Al$_2$O$_3$ as shown in
figure~\ref{fig-1} at different magnitudes of the beam
current.

\begin{figure}[h]
\centerline{\includegraphics[width=0.51\textwidth]{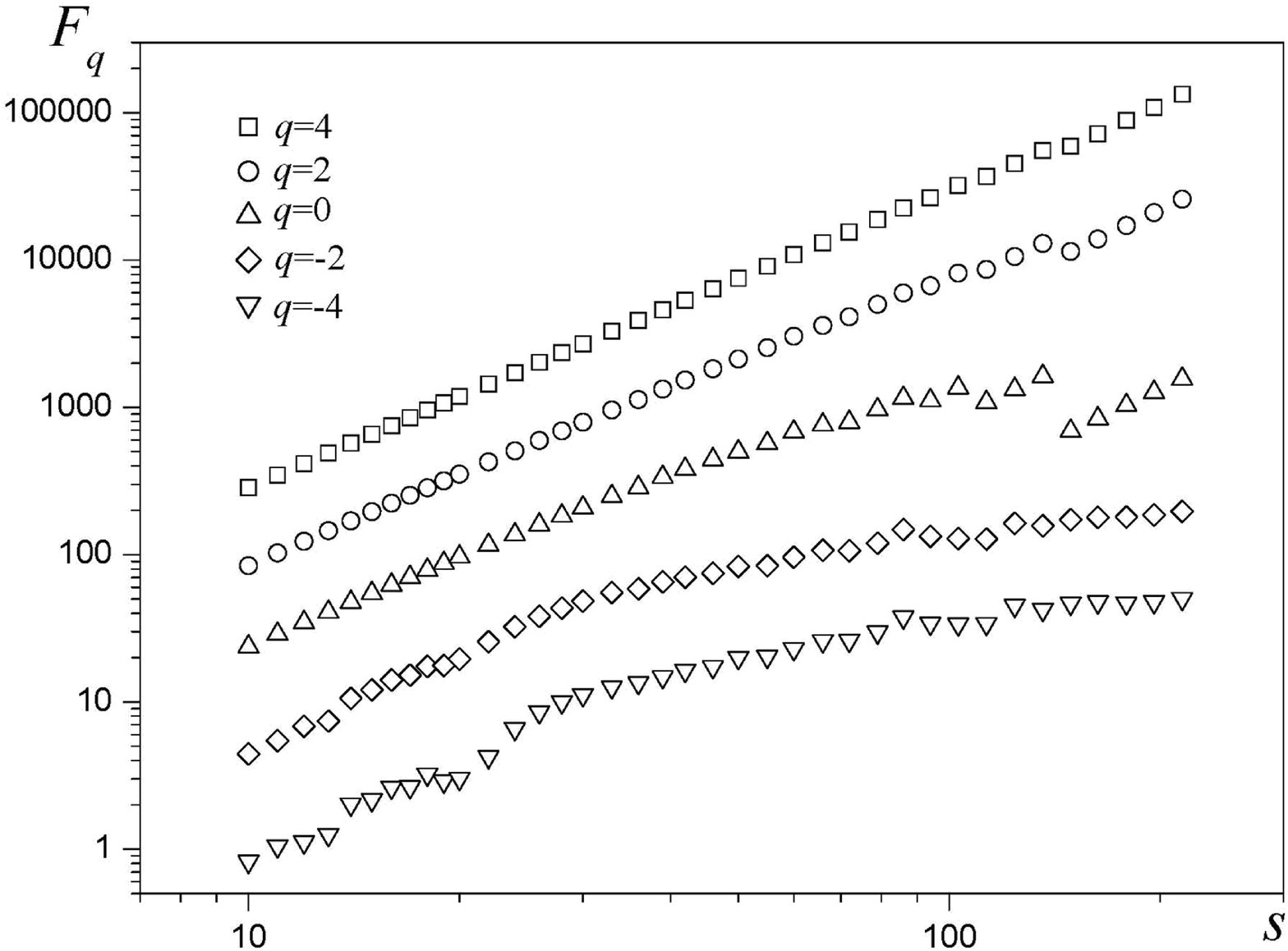}}%
\caption{Log-log plot of the fluctuation function $F_{q}(s)$ versus
the scale $s$ for five different values of $q$, calculated for SEM
image of the surface of hybrid coating, modified by the beam current
$I=20$~mA.} \label{fig-2}%
\end{figure}
\begin{figure}[h]
\includegraphics[height=5cm]{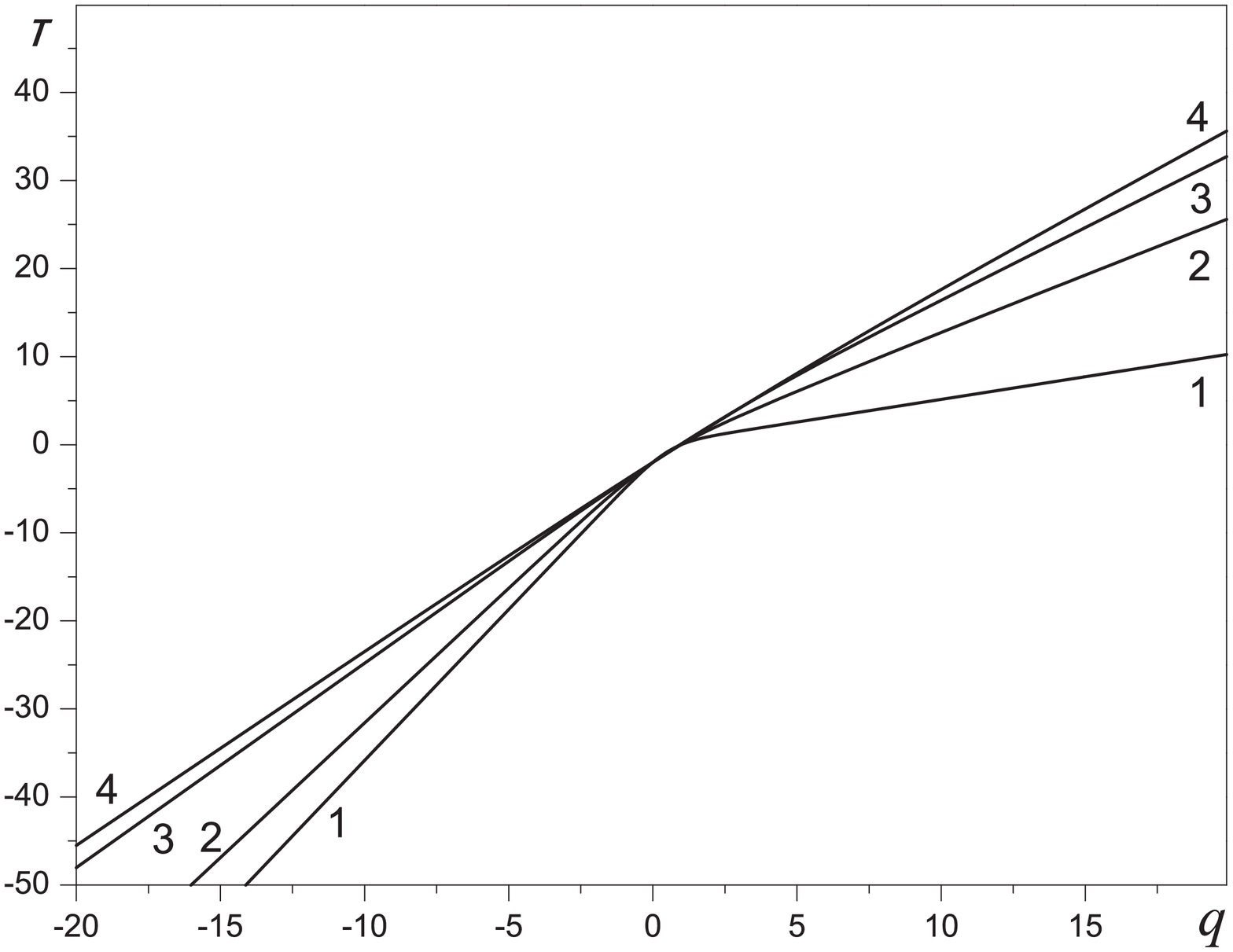}%
\hfill%
\includegraphics[height=5.1cm]{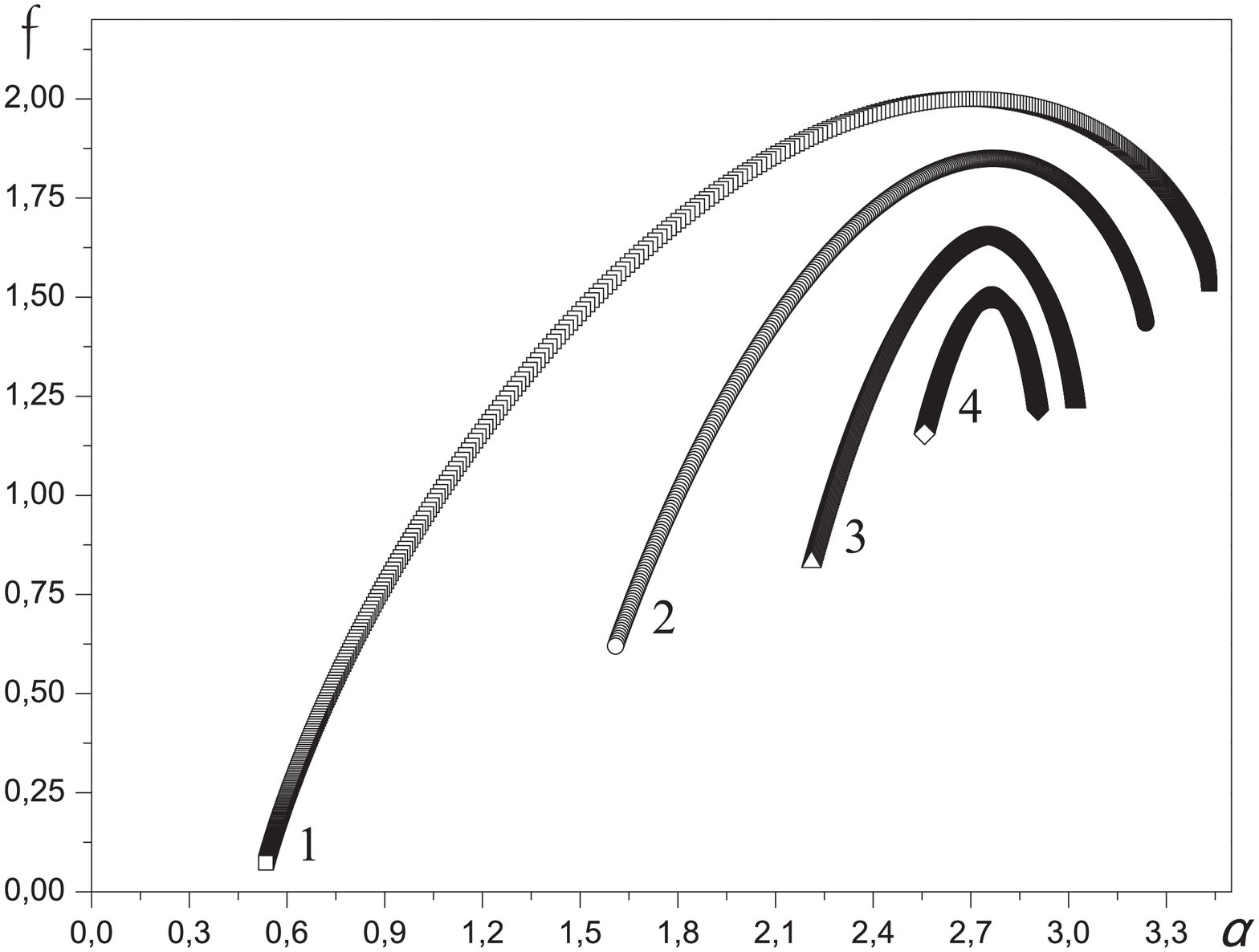}%
\hspace{5mm}
\\%
\parbox[t]{0.48\textwidth}{%
\caption{Mass exponent $\tau(q)$ as a function of $q$. Lines 1--4
correspond to the surfaces modificated by HCEB with current $20$~mA,
$20+15$~mA, $20+25$~mA, $20+35$~mA, respectively.} \label{fig-3}%
}%
\hfill%
\parbox[t]{0.48\textwidth}{%
\caption{Multifractal spectra of  hybrid coating surfaces
TiN/Al$_{2}$O$_{3}$ under HCEB modification. Lines 1--4 correspond to
the surfaces  modified by HCEB with current $20$~mA, $20+15$~mA,
$20+25$~mA, $20+35$~mA, respectively. The curves 2--4 have been shifted
for clarity.} \label{fig-4}%
}%
\end{figure}

If the object under investigation has a self-similar structure,
relation (\ref{eq7}) is expected to be linear in double logarithmic
scales. Figure~\ref{fig-2} illustrates the dependence of the
fluctuation function $F_q(s)$ on the scale $s$ for different values
of $q$, calculated for TiN/Al$_2$O$_3$ surface modified by the beam
current $I=20$~mA. As it follows from figure~\ref{fig-2}, the
dependence $F_{q}(s)$ has a clear linear part, which means that the
surface of a hybrid coating has a self-similar structure. On the other
hand, in the range where $q<0$, the calculation is expected to yield
a large error.

We have computed the mass exponent $\tau(q)$ for different surfaces
in the range $-20<q<20$. Figure~\ref{fig-3} shows $\tau(q)$ as a
function of $q$ for hybrid coating surfaces under HCEB modification
at different current magnitudes. The nonlinearity of $\tau(q)$
indicates that the surface has a multifractal structure, i.e., it cannot be completely described by a single value of a fractal dimension
$\alpha$. Different values of $\alpha$ are related to the segments of the
surface with  different values of the fluctuation function $F_{q}$.
The latter is calculated as the difference of the surface local heights from
some smooth fitting function. Thus, the set of the fractal exponents
$f(\alpha)$ can be considered as a quantitative measure of the surface
roughness. The strongest multifractality was observed for the surface
being modified with the beam current magnitude $I=20$~mA, becoming more
weaker with the growth of the beam current. It is shown that
geometry of the surface layers of hybrid coatings depends on the
electron density of the beam power.

We have calculated the values of the singularity strength function
$\alpha$ and the multifractal spectrum $f(\alpha)$  using equations
(\ref{eq10}) and (\ref{eq11}). Figure~\ref{fig-4} shows the spectrum
$f(\alpha)$ for four samples under investigation. As it can be seen,
the width of $f(\alpha)$ is different for the samples treated with
different density of the beam current. The more uniform is the surface, the
more restricted is the spectrum $f(\alpha)$.

Minimum and maximum
values of $\alpha (q)$  are important statistical parameters that describe the
multifractal nature of fracture surfaces. These values are the singularity strengths
associated with the region of the sets where the measures are the
least and the most singular, respectively \cite {Li09}.

In the formalism of multifractals, $\alpha_\textrm{min}$ is related to the
maximum probability measure by
$P_\textrm{max}\sim\varepsilon^{-\vartriangle\alpha}$, where $\varepsilon$
represents the scale approaching zero and it is a small quantity,
whereas $\alpha_\textrm{max}$ is related to the minimum probability measure
through $P_\textrm{min}\sim\varepsilon^{\alpha_\textrm{max}}$. The width
$\vartriangle\alpha$ can be used to describe the range of the
probability measures \cite {Li09}:
\begin{equation}
\label{eq12}
\frac{P_\textrm{max}}{P_\textrm{min}}\sim\varepsilon^{-\vartriangle\alpha}\,.
\end{equation}

The larger is $\alpha$, the wider is the probability
distribution, and the strongest is the difference between the highest and the
lowest growth probability.
\begin{figure}[h]
\centerline{\includegraphics[height=5cm]{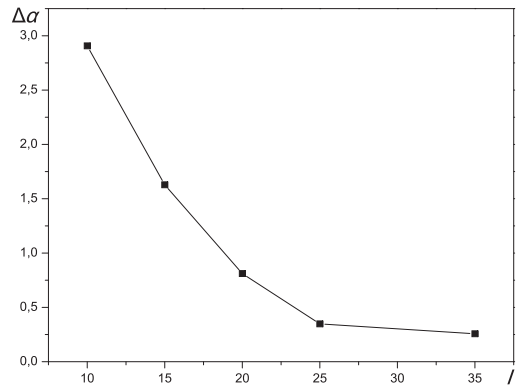}}
\caption{The width of multifractal spectrum $\vartriangle\alpha$ for
surfaces being modified with different beam current magnitudes.}
\label{fig-5}%
\end{figure}

\begin{figure}[h]
\centerline{\includegraphics[height=5.6cm]{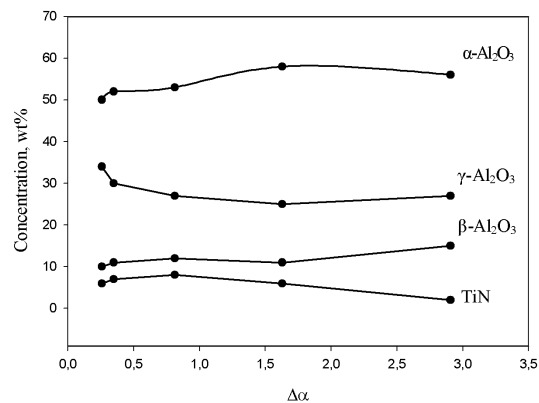}}%
\caption{The dependence of phase composition for hybrid coatings after
duplex melting of their surfaces with different singularity
strengths.} \label{fig-6}%
\end{figure}

Figure~\ref{fig-5} illustrates the relation between the width of the
multifractal spectrum and  different values of the beam current. As
it can be seen, the width of the multifractal spectrum decreases with
an increase of the current magnitude. The change of the surface relief,
as shown in figure~\ref{fig-5}, proves the correspondence between
the theoretical calculations and experimental results, i.e., with an
increase of the beam current, the surface becomes more regular. A
significant reduction of multifractal spectrum width with the
increase of the current magnitude of 25~mA occurs due to remelting
which smooths out the craters produced by  degassing. A further
increase of the beam current provides only an enhanced homogeneity of
the structure and mass transfer processes between the layers
composing the coating matrix. Figure~\ref{fig-6} shows the
dependence of phase composition  on the singularity strength for hybrid coatings after duplex melting of their surfaces.

Special attention was paid to the change of concentration of the elements in
the hybrid coating after duplex melting. It was shown in
work \cite{Pog06} that the initial composition was
about 60 wt. \% of Al$_{2}$O$_{3}$. All the other phases and compounds
comprised 40 wt. \%.

As we can see from figure~\ref{fig-6}, with an increase of the
density of the energy flow and, consequently, a decrease of the
spectrum width, $\triangle \alpha$ was accompanied by an
insignificant increase in $\alpha$-Al$_{2}$O$_{3}$ amount. A further
increase of the current led to a decrease in the percentage content of
$\alpha$-phase from 58 wt. \% to 50 wt. \% and to an increase of
$\gamma$-Al$_{2}$O$_{3}$ from 25 wt. \% to 34 wt. \%. The
concentration of other phases and compounds underwent
insignificant changes.  Hence, a non-monotonous change of the
phase percentage content was associated with an increase of the
energy density. As a result, the surface of hybrid coating became
more uniform by TiN and Al$_{2}$O$_{3}$ coating melting.

\section{Conclusion}

The mechanical studies demonstrated that hybrid coatings
based on corundum and titanium nitride, which were modified  by an
electron beam until melted, possessed notably better servicing
characteristics. Therefore, this technology could be applied to solve
technical problems (for example, to decrease wear, to protect from
corrosion, to increase adhesion and to improve nano- and
micro-hardness).

Quantitative parameters of the surface structure obtained by the
two-dimensional multifractal fluctuation method can be used to
characterize the topology of the interface under modification. As
shown by the numerical analysis, the character of surface morphology
changed from high non-uniform roughness to smoothed regions with a
gradual increase of the current density electron beam.

\newpage
\ukrainianpart

\title{Чисельний аналіз морфологічних та фазових перетворень покриття TiN/Al$_{2}$O$_{3}$ під час модифікації низькоенергетичним сильнострумовим електронним пучком}
\author{О.Д. Погребняк, В.М. Борисюк, А.А. Багдасарян}
\address{Сумський державний унiверситет , вул. Римського-Корсакова, 2, 40007 Суми, Україна}

\makeukrtitle

\begin{abstract}
\tolerance=3000%
Проаналізовано процес модифікації структури поверхні гібридного
покриття TiN/Al$_{2}$O$_{3}$ під впливом низькоенергетичного
сильнострумового електронного пучка. Шорсткість поверхні розглянуто
як функцію струму пучка. Поверхні отриманих зразків досліджувались
за допомогою двовимірного мультифрактального флуктуаційного аналізу.
Для кількісного аналізу зміни шорсткості розрахована функція
мультифрактального спектру. Показано, що зі збільшенням енергії
пучка поверхня стає більш регулярною та рівномірною.
\keywords самоподібність, фрактальна розмірність, гібридне покриття,
НПЕП ефект

\end{abstract}


\begin{thebibliography}{99}

\bibitem{Fed98} Feder~J., Fractals,
    Plenum Press, New-York, London, 1998.

\bibitem{Ole09} Olemskoi~A., Danylchenko~S., Borisyuk~V., Shuda~I.,
    Metallofiz. Noveishie Tehknol., 2009, \textbf{31}, 777.

\bibitem{Ol95} Olemskoi~A.A., Fractal in Condensed Matter Physics,
    In: Physics Reviews, 1995, \textbf{18}, Part~1, 1.

\bibitem{Ol09} Olemskoi~A.A., Synergetics of Complex Systems: Phenomenology and Statistical Theory, KRASAND, Moscow, 2009 (in Russian).

\bibitem{Arn00} Arn\'{e}odo~A., Decoster~N.,
    Roux~S.G., Eur. Phys. J. B, 2000, \textbf{15}, 567; \doi{10.1007/s100510051161}.

\bibitem{Jen03} Jeng~Y.R., Tsai~P.S., Fang~T.H., Microelectron.
    Eng., 2003, \textbf{65}, 406; \doi{10.1016/S0167-9317(03)00052-2}.

\bibitem{Pfe89} Pfeifer~P.,  Wu~Y.J., Cole~M.W., Krim~J.,
    Phys. Rev. Lett., 1989, \textbf{62}, 1997; \doi{10.1103/PhysRevLett.62.1997}.

\bibitem {Bor20} Borisyuk~V.N., Kassi~J., Holovchenko~A.I.,
    J. Nano-Electron. Phys., 2011, \textbf{3},  No.~4, 20.

\bibitem{Gu06} Gu~G.F.,  Zhou~W.X.,
    Phys. Rev. E, 2006, \textbf{74}, 061104; \doi{10.1103/PhysRevE.74.061104}.

\bibitem{Li09} Liu~C., Jiang~X.L., Liu~T., Zhao~L., Zhou~W.X., Yuan~W.K.,
     Appl. Surf. Sci., 2009, \textbf{255}, 4239; \doi{10.1016/j.apsusc.2008.11.014}.

\bibitem{Niu08} Niu~M.R., Zhou~W.X., Yan~Z.Y., Guo~Q.H., Liang~Q.F.,
     Wang~F.C., Yu~Z.H., Chem. Eng. J., 2008, \textbf{143}, 230; \doi{10.1016/j.cej.2008.04.011}.

\bibitem{Hur51} Hurst~H.E., Trans. Am. Soc. Civ. Eng., 1951, \textbf{116}, 770.

\bibitem{Ol20} Olemskoi~A., Shuda~I., Borisyuk~V., Europhys. Lett., 2010, \textbf{89}, 50007; \doi{10.1209/0295-5075/89/50007}.

\bibitem{Kan02} Kanthelhardt~J.W., Zscheinger~S.A.,
    Koscienly-Bunde~E., Havlin~S., Bunde~A., Stanley~H.E.,
    Physica A, 2002, \textbf{316}, 87; \doi{10.1016/S0378-4371(02)01383-3}.

\bibitem{Bag12} Pogrebnyak~A.D., Borisyuk~V.N., Bagdasaryan~A.A., In:
    Proceedings of the International Conference ``Nanomaterials: Applications and Properties'' (Alushta, 2012), Vol.~1, 2012, 02NFC27.

\bibitem{Pog06} Pogrebnyak~A.D., Kravchenko~Yu.A., Kislitsyn~S.M.,
    Ruzimov~Sh.M., Noli~F., Misaelides~P., Hatzidimitriou~A., Surf.
    Coat. Technol., 2006, \textbf{201}, 2621; \doi{10.1016/j.surfcoat.2006.05.018}.

\bibitem {Pog12} Pogrebnjak~A. D., Ponomarev~A.G., Shpak~A.P.,
    Kunitskii~Yu.A., Phys. Usp., 2012, \textbf{55}, 270; \doi{10.3367/UFNe.0182.201203d.0287}.

\bibitem {Pog20} Pogrebnyak~A.D., Sobol~O.V., Beresnev~V.M., Turbin~P.V.,
    Il'yashenko~M.V., Kirik~G.V., Makhmudov~N.A., Shypylenko~A.P.,
    Kaverin~M.V., Tashmetov~M.Yu., Pshyk~A.V., In: Nanostructured Materials
    and Nanotechnology IV: Ceramic Engineering and Science Proceedings,
    2010, \textbf{31}, 127; \doi{10.1002/9780470944042.ch14}.

\bibitem{Kun07} Kunchenko~Yu.V., Kunchenko~V.V., Nekliudov~I.M., Kartmazov~G.N., Andreev~A.A.,  Probl. Atomic Sci. Technol., 2007, \textbf{2}, 203.

\bibitem{Pei92} Peitgen~H., J\"{u}rgens~H., Saupe~D., Chaos and Fractals: New
    Frontiers of Sciences, Springer-Verlag, New-York, 1992.

\bibitem{Hal86} Halsey~T.C., Jensen~M.H., Kadanoff~L.P.,  Procaccia I., Shraiman~B.I., Phys. Rev. A, 1986, \textbf{33}, 1141; \doi{10.1103/PhysRevA.33.1141}.

    \end{thebibliography}
\end{document}